\documentstyle[12pt]{article}

\begin{document}
\topmargin 0pt
\oddsidemargin 0mm
\vspace{3mm}
\begin{flushright} JLAB-THY-97-31 \end{flushright}
\vspace{10mm}
\begin{center}
{\Large{\bf How to extract the $P_{33}(1232)$ resonance contributions
from the amplitudes
$M_{1+}^{3/2},E_{1+}^{3/2},S_{1+}^{3/2}$
of pion electroproduction on nucleons}}\\
\vspace{1cm}
{\large I.G.Aznauryan}\\
\vspace{1cm}
{\em Yerevan Physics Institute}\\
{Alikhanian Brothers St.2, Yerevan, 375036 Armenia}\\
{(e-mail addresses: aznaury@vx1.yerphi.am, aznaury@cebaf.gov)}\\
\vspace{5mm}
Abstract\\
\vspace{3mm}
\end{center}
Within the dispersion relation approach,
solutions of integral equations for the multipoles
$M_{1+}^{3/2},E_{1+}^{3/2},S_{1+}^{3/2}$
are found at
$0\leq Q^2\leq 3~GeV^2$. These solutions
should be used as input for the
resonance and nonresonance
contributions in the analyses  of pion electroproduction
data in the $P_{33}(1232)$ resonance region.
It is shown that the traditional identification
of the amplitude 
$M_{1+}^{3/2}$ (as well of the amplitudes $E_{1+}^{3/2},S_{1+}^{3/2}$)
with the $P_{33}(1232)$ resonance contribution is not right;
there is a contribution in these amplitudes which has a nonresonance nature
and is produced by rescattering effects in the diagrams
corresponding to the nucleon and pion poles.
This contribution is reproduced by the dispersion relations.
Taking into account nonresonance contributions
in the amplitudes
$M_{1+}^{3/2},E_{1+}^{3/2}$,
the helicity amplitudes $A_p^{1/2},~A_p^{3/2}$ and the ratio
$E2/M1$ for the
$\gamma N \rightarrow P_{33}(1232)$
transition are extracted from experiment at $Q^2=0$.
They are in good agreement with 
quark model predictions.
\vspace{5mm}
\newline
PACS number(s): 11.55.Fv, 11.80.Et, 13.60.Le, 25.20.Lj, 25.30Rw 
\vspace{5mm}
\renewcommand{\thefootnote}{\arabic {footnote}}
\setcounter{page}{1}
\section{Introduction}
It is known that experimental data
on form factors of the
$\gamma N \rightarrow P_{33}(1232)$
transition may play an important role
in the investigation of energetic scale of transition
to perturbative region of QCD.
The conservation of quark helicities in the regime 
of perturbative QCD
leads to the asymptotic relation [1-3]:
\begin{equation} 
\frac{G_E}{G_M}\rightarrow -1,~~Q^2\rightarrow \infty~~(pQCD).
\end {equation}

In contrast to this
at $Q^2=0$ quark model predicts the suppression
of $G_E$:
\begin{equation} 
\frac{G_E}{G_M}=0,~~Q^2=0~~(quark~model),
\end {equation}
which agrees well with experiment.
Thus, the transition from nonperturbative
region of QCD to perturbative one
is characterized by a
striking change of the behavior of the ratio
$G_E(Q^2)/G_M(Q^2)$ , and, therefore,
the measurement of this ratio
will provide a sensitive test for
understanding of mechanisms of transition
to the QCD asymptotics.

For the Coulombic form factor
predictions of quark model and pQCD
coincide with each other:
\begin{eqnarray} 
&&\frac{G_C}{G_M}=0,~~Q^2=0~~(quark~model),\\
&&\frac{G_C}{G_M}\rightarrow 0,~~Q^2\rightarrow \infty~~(pQCD),
\end {eqnarray}
and in this case we have no test 
for investigation of the energetic scale
of the transition to the QCD asymptotics.
However, precise measurement of the $Q^2$-dependence
of the Coulombic form factor can be of interest
too for the development of realistic models
of the nucleon and the $P_{33}(1232)$.

In the nearest future significant
progress in the investigation of the
$\gamma N \rightarrow P_{33}(1232)$
transition form factors is expected due to 
construction of continuous wave
electron accelerators. These form factors
will be studied in the reaction of pion
electroproduction on nucleons
via extraction of the resonance multipole
amplitudes
$M_{1+}^{3/2},E_{1+}^{3/2},S_{1+}^{3/2}$
which carry information
on the contribution of the resonance
$P_{33}$ to this process
(the diagram of Fig. 1d).

It is well known that an extremely
fruitful role in the investigation of pion
photo-and electroproduction on nucleons belongs
to the approach based on dispersion relations.
The basis of this approach was founded
in the classical works [4,5].
Further, it was developed
in numerous works, among which
let us mention the papers [6-12].
Analysing the results obtained
within dispersion relation approach 
we came to the conclusion that in order
to obtain a proper input for the analysis 
of expected experimental data
in the $P_{33}(1232)$ resonance
region it is very useful to use the approach
developed in Refs. [6,7]. Let us clarify
this statement.

The solutions of integral equations for the multipoles
$M_{1+}^{3/2},E_{1+}^{3/2},S_{1+}^{3/2}$
following from dispersion relations
for these multipoles are obtained
in Refs. [6,7] in the form
which contains two parts.
One part is the particular solution
of the integral equation generated by the Born term
(i.e. by the diagrams of Fig. 1(a-c) corresponding
to the nucleon and pion poles).
It has definite magnitude fixed by the Born term.
Our analysis shows that
other contributions to the particular solutions
for the multipoles  
$M_{1+}^{3/2},E_{1+}^{3/2},S_{1+}^{3/2}$
which can arise from nonresonance multipoles
and high energy contributions
are negligibly small.

Other part of the solutions
corresponds to the homogeneous parts
of the integral equations.
This part has a certain
energy dependence fixed by dispersion relations
and an arbitrary weight which should be
found from the comparison with experiment.

In the present work (Sec.3) we have repeated
the results of Refs. [6,7]
for the solutions of dispersion relations
for the multipoles
$M_{1+}^{3/2},E_{1+}^{3/2}$
at $Q^2=0$. In addition, we have obtained
$Q^2$-evolution of these solutions 
in the range of $Q^2$
from 0 to $3~GeV^2$.
The solutions of dispersion
relations for the multipole
$S_{1+}^{3/2}$ 
at $0\leq Q^2\leq 3~GeV^2$
are also obtained. The obtained solutions should be
considered as an input for the multipoles
$M_{1+}^{3/2},E_{1+}^{3/2},S_{1+}^{3/2}$
in the analysis of
future experimental data.

Analysing the solutions of integral equtions
for multipole amplitudes obtained within approach
of Refs. [6,7] we came also to the conclusion
that the traditional identification
of the amplitude $M_{1+}^{3/2}$
(as well of the amplitudes $E_{1+}^{3/2},S_{1+}^{3/2}$)
with the 
contribution of the $P_{33}(1232)$ resonance
is not right. The physical interpretation of
these solutions (Sec. 3) shows that the particular solutions 
of integral equations generated by the Born term
should be considered as nonresonance background 
to the contribution of the $P_{33}$ resonance.
These solutions are produced by
rescattering effects in the diagrams 1(a-c)
corresponding to the 
Born term. For the first time
the presence of such nonresonance contribution in the
multipoles $M_{1+}^{3/2},E_{1+}^{3/2}$
was mentioned in Ref. [13] within dynamical model
describing pion photoproduction in the $P_{33}$ 
region in terms of the diagrams of Fig. 1 taking into 
account rescattering effects.

In Sec. 4 taking into account nonresonance 
contributions in the multipoles
$M_{1+}^{3/2},E_{1+}^{3/2}$
the amplitudes
$A_{1/2}^p,~A_{3/2}^p$ for the
$\gamma N\rightarrow P_{33}(1232)$ 
transition are extracted from experiment.
The magnitudes of these amplitudes
turned out to be smaller than the magnitudes
which are traditionally extracted from experiment 
without taking into account nonresonance background
contributions in the amplitude
$M_{1+}^{3/2}$. As a result, the traditionally mentioned disagreement
between quark model predictions and experiment
for the amplitudes
$A_{1/2}^p,~A_{3/2}^p$ turned out to be removed.
\section{Dispersion relations for
invariant and multipole amplitudes}
In this Section we will present very briefly
main formulas which are necessary for our calculations.
Following the work [11] we choose invariant amplitudes
in accordance with the following
definition of the hadron current:
\begin{eqnarray}
I^\mu =\bar{u}(p_2)\gamma _5 \left\{ \frac{B_1}{2}\left[ \gamma
^\mu (\gamma k)-(\gamma k)\gamma ^\mu \right]+2P ^\mu
B_2+2q^\mu B_3+2k^\mu B_4\right.\nonumber \\
\left.-\gamma ^\mu B_5+(\gamma k)P^\mu B_6+
(\gamma k)k^\mu B_7+(\gamma k)q^\mu B_8\right\} u(p_1),
\end{eqnarray}
where $k,q,p_1,p_2$ are the 4-momenta
of virtual photon, pion, initial and final nucleons, respectively,
$P=\frac{1}{2}(p_1+p_2),~Q^2\equiv-k^2$, $B_1,B_2,...B_8$
are invariant amplitudes which are functions
of the invariant variables $s=(k+p_1)^2,~t=(k-q)^2,~Q^2$.

The conservation of the hadron current
leads to the relations:
\begin{eqnarray}
&&4Q^2B_4=(s-u)B_2 -2(t+Q^2-\mu ^2)B_3, \\
&&2Q^2B_7=-B'_5 -(t+Q^2-\mu ^2)B_8,
\end{eqnarray}
where
$B'_5\equiv B_5-\frac{1}{4}(s-u)B_6$
$\mu$ is the pion mass.
So, only the six of the eight invariant amplitudes
are independent. Let us choose as
independent amplitudes following ones:
$B_1,B_2,B_3,B'_5,B_6,B_8$.
For all these amplitudes, except $B_3^{(-)}$,
unsubtracted dispersion relations can be written:
\begin{eqnarray}
Re B_i^{(\pm,0)}(s,t,Q^2)=&&R_i^{(v,s)}
\left(\frac{1}{s-m^2}\pm \frac{\eta_i }{u-m^2}\right)\nonumber\\
&&+\frac{P}{\pi }\int \limits_{s_{thr}}^{\infty}
Im B_i^{(\pm,0)}(s',t,Q^2)
\left(\frac{1}{s'-s}\pm \frac{\eta_i }{s'-u}\right) ds',
\end{eqnarray}
where $\pm$ and $0$ labels refer to isospin states
with a definite symmetry under the
interchange $s\leftrightarrow u$,
$R_i^{(v,s)}$ are residues in 
the nucleon poles (they are given in the Appendix),
$\eta_1=\eta_2=\eta_6=1,~\eta_3=\eta'_5=\eta_6=-1$,
$s_{thr}=(m+\mu)^2$, m is
the nucleon mass.
For the amplitude $B_3^{(-)}$
we take the subtraction point at an infinity.
In this case using the current conservation
condition (6) we have
\begin{eqnarray}
Re B_3^{(-)}(s,t,Q^2)=&&R_3^{(v)}
\left(\frac{1}{s-m^2}+\frac{1}{u-m^2}\right)-
\frac{eg}{t-\mu ^2 }F_\pi (Q^2)\nonumber\\
&&+\frac{P}{\pi } \int \limits_{s_{thr}}^{\infty}
Im B_3^{(-)}(s',t,Q^2)
\left(\frac{1}{s'-s}+\frac{1 }{s'-u}-\frac{4 }{s'-u'}\right)ds'.
\end{eqnarray}
In order to connect the invariant
amplitudes with cross section, helicity
and multipole amplitudes, it is convenient to
introduce intermediate amplitudes
$f_i$ which are related
to the invariant amplitudes via:
\begin{eqnarray}
f_1=&&\frac{a_1}{8\pi W}\left[(W-m)B_1-B_5\right], \\
f_2=&&\frac{a_2}{8\pi W}\left[-(W+m)B_1-B_5\right], \\
f_3=&&\frac{a_3}{8\pi W}
\left[2B_3-B_2+(W+m)\left(\frac{B_6}{2}-B_8 \right)\right],\\
f_4=&&\frac{a_4}{8\pi W}
\left[-(2B_3-B_2)+(W-m)\left(\frac{B_6}{2}-B_8\right)\right],\\
f_5=&&\frac{a_5}{8\pi W}
\left\{\left[Q^2B_1+(W-m)B_5+2W(E_1-m)\left(B_2-\frac{W+m}{2}B_6\right)
\right](E_1+m)\right.\nonumber \\
&&\left.-X\left[(2B_3-B_2)+(W+m)\left(\frac{B_6}{2}- B_8 \right)\right]\frac{}{}\right\}, \\
f_6=&&\frac{a_6}{8\pi W}
\left\{-\left[Q^2B_1-(W+m)B_5+2W(E_1+m)\left(B_2+\frac{W-m}{2}B_6
\right)\right](E_1-m)\right.\nonumber \\
&&\left.+X\left[(2B_3-B_2)-(W-m)\left(\frac{B_6}{2}- B_8 \right) \right]\frac{}{}\right\},\\
\end{eqnarray}
where
\begin{eqnarray}
&&a_1=\left[(E_1+m)(E_2+m)\right]^{1/2},\nonumber \\
&&a_2=\left[(E_1-m)(E_2-m)\right]^{1/2},\nonumber \\
&&a_3=\left[(E_1-m)(E_2-m)\right]^{1/2} (E_2+m), \\
&&a_4=\left[(E_1+m)(E_2+m)\right]^{1/2}(E_2-m),\nonumber \\
&&a_5=\left[(E_1-m)(E_2+m)\right]/Q^2,\nonumber \\
&&a_6=\left[(E_1+m)(E_2-m)\right]/Q^2,\nonumber
\end{eqnarray}
and
\begin{equation}
X=\frac{k_0}{2}(t-\mu ^2+Q^2)-Q^2q_0,
\end{equation}
$k_0,q_0,E_1,E_2$ are the energies
of virtual photon, pion, initial and final nucleons
in the c.m.s., $W=s^{1/2}$.
The amplitudes $f_i$ are related
to cross section, helicity and multipole
amplitudes by the relations (A.3-A.5).

Dispersion relations for the multipoles
$M_{1+},E_{1+},S_{1+}$ can be found from the
relations (8,9)
using the projection formulas:
\begin{eqnarray}
&&M_{1+}=\frac{1}{8}\int\limits_{-1}^{1}\left[2f_1x+f_2(1-3x^2)-f_3
(1-x^2)\right]dx,\nonumber \\
&&E_{1+}=\frac{1}{8}\int\limits_{-1}^{1}\left[2f_1x+f_2(1-3x^2)+(f_3
+2x f_4)(1-x^2)\right]dx, \\
&&S_{1+}=\frac{1}{8}\int\limits_{-1}^{1}\left[2f_5 x+(3x^2-1)f_6\right]
dx.\nonumber
\end{eqnarray}

All the subsequent calculations for these multipoles
will be made numerically.
By this reason we do not specify further these relations,
and dispersion relations for
$M_{1+},E_{1+},S_{1+}$ will be written
immediately through the dispersion
relations (8,9).

\section{Solutions of dispersion relations
for the multipoles 
$M_{1+}^{3/2},E_{1+}^{3/2},S_{1+}^{3/2}$
at $Q^2\leq3~GeV^2$. The interpretation
of these solutions}
Let us write the dispersion relations for the multipoles
$M_{1+}^{3/2},E_{1+}^{3/2},S_{1+}^{3/2}$
in the form:
\begin{eqnarray}
M(W,Q^2)=&&M^B(W,Q^2)+
\frac{1}{\pi}\int\limits_{W_{thr}}^{\infty}
\frac{ImM(W',Q^2)}{W'-W-i\varepsilon}dW'\nonumber\\
&&+\frac{1}{\pi}\int\limits_{W_{thr}}^{\infty}
K(W,W',Q^2)ImM(W',Q^2)dW'.
\end{eqnarray}
Here $M(W,Q^2)$ denotes any of the considered
multipoles, $M^B(W,Q^2)$ is the contribution
of the Born term into these multipoles,
$K(W,W',Q^2)$ is a nonsingular kernel arising
from the u-channel contribution into the
dispersion integral and the nonsingular part
of the s-channel contribution.
In the integrand of the relation (20) we did not take
into account the coupling of 
$M(W,Q^2)$ to other multipoles by the following reason.
Here we consider only resonance multipoles
with large imaginary parts.
Contributions of nonresonance multipoles
to these multipoles are negligibly small.
The couplings of the resonance 
multipoles with each other are also reasonably small [7].
In the dispersion relation (20) we also did not take into
account high energy contributions
to the multipoles 
$M(W,Q^2)$ which, if they exist,
should be added to $M^B(W,Q^2)$.
Our estimations show that these
contributions to the multipoles
$M_{1+}^{3/2},E_{1+}^{3/2},S_{1+}^{3/2}$
are negligibly small and do not affect
our results presented below.

In the present work we 
use the dispersion relations in the
$P_{33} (1232)$ resonance region. From the phase shift
analyses of $\pi N$ scattering in this region [14-17]
it is known that the resonance amplitude 
$h_{1+}^{\frac{3}{2}}(W)$
of $\pi N$ scattering is elastic, and, so,
can be written in the form:
\begin{equation}
h_{1+}^{\frac{3}{2}}(W)=
sin \delta_{1+}^{\frac{3}{2}}(W)
exp(i\delta_{1+}^{\frac{3}{2}}(W)).
\end{equation}

In this energy region due to the elasticity
of the amplitudes $(1+)$ one can use the Watson theorem [18]
for the amplitudes $M(W,Q^2)$ and write them in the form: 
\begin{equation}
M(W,Q^2)=
exp(i\delta_{1+}^{\frac{3}{2}}(W))
|M(W,Q^2)|.
\end{equation}

From the experimental data [14-17] it is known
that for the amplitude $h_{1+}^{\frac{3}{2}}(W)$
the elasticity condition is strictly fulfilled
up to $W=1.5~GeV$, i.e. up to energies which
are much higher in comparison with the
energies in the $P_{33} (1232)$ resonance region.
By this reason one can assume that the amplitude
$h_{1+}^{\frac{3}{2}}(W)$
is elastic at all energies and the condition (22)
is valid on the whole physical cut, i.e.
\begin{equation}
ImM(W,Q^2)=h^* (W)M(W,Q^2).
\end{equation}

Furthemore, as at $W=1.5~GeV,~\delta (W)=160^{\circ}$,
one can assume that
\begin{equation}
0<\delta (\infty)\leq \pi.
\end{equation}

With these conditions the dispersion relation (20)
transforms into singular integral equation.
At $K(W,W',Q^2)=0$ this equation 
has a solution
which have the following analytical form
(see Ref.[6] and the references therein):
\begin{equation}
M_{K=0}(W,Q^2)=M^{part}_{K=0}(W,Q^2)+c_M M^{hom}_{K=0}(W),
\end{equation}
where
\begin{equation}
M^{part}_{K=0}(W,Q^2)=M^B(W,Q^2)+
\frac{1}{\pi}\frac{1}{D(W)}
\int\limits_{W_{thr}}^{\infty}
\frac{D(W')h(W')M^B(W',Q^2)}{W'-W-i\varepsilon}dW'
\end{equation}
is the particular solution of the singular equation, and
\begin{equation}
M^{hom}_{K=0}(W)=\frac{1}{D(W)}=
exp\left[\frac{W}{\pi}
\int\limits_{W_{thr}}^{\infty}
\frac{\delta (W')}{W'(W'-W-i\varepsilon)}dW'\right]
\end{equation}
is the solution of the homogeneous equation
\begin{equation}
M^{hom}_{K=0}(W)=
\frac{1}{\pi}
\int\limits_{W_{thr}}^{\infty}
\frac{h^* (W')M^{hom}_{K=0}(W')}{W'-W-i\varepsilon}dW',
\end{equation}
which enters the solution (25) with an arbitrary weight,
i.e. multiplied by an arbitrary constant $c_M$.

Replacing in the Eqs. (25,26) $M^B(W,Q^2)$ by
\begin{equation}
M^B(W,Q^2)+
\frac{1}{\pi}
\int\limits_{W_{thr}}^{\infty}
K(W,W',Q^2)h^*(W')M(W',Q^2)dW'\nonumber,
\end{equation}
one can transform the dispersion relation (20)
into the Fredholm integral equation for the imaginary parts
of the multipole amplitudes:
\begin{eqnarray}
ImM(W,Q^2)=&&ImM^{part}_{K=0}(W,Q^2)+c_M ImM^{hom}_{K=0}(W)+\nonumber\\
&&\frac{1}{\pi}
\int\limits_{W_{thr}}^{\infty}
f(W,W',Q^2)ImM(W',Q^2)dW,'
\end{eqnarray}
where
\begin{equation}
ImM^{part}_{K=0}(W,Q^2)=sin \delta (W)\left[M^B(W,Q^2)cos \delta (W)+
e^{a(W)}r(W,Q^2)\right],
\end{equation}
\begin{equation}
r(W,Q^2)=
\frac{P}{\pi}
\int\limits_{W_{thr}}^{\infty}
\frac{e^{-a(W')}sin \delta (W')M^B(W',Q^2)}{W'-W}dW',
\end{equation}
\begin{equation}
ImM^{hom}_{K=0}(W)=sin \delta (W)
e^{a(W)},
\end{equation}
\begin{equation}
f(W,W',Q^2)=sin \delta (W)\left[K(W,W',Q^2)cos \delta (W)+
e^{a(W)}R(W,W',Q^2)\right],
\end{equation}
\begin{equation}
R(W,W',Q^2)=
\frac{P}{\pi}
\int\limits_{W_{thr}}^{\infty}
\frac{e^{-a(W'')}sin \delta (W'')K(W'',W',Q^2)}{W''-W}dW'',
\end{equation}
\begin{equation}
a(W)=
\frac{P}{\pi}
\int\limits_{W_{thr}}^{\infty}
\frac{W \delta (W')}{W'(W'-W)}dW'.
\end{equation}

The solution of Eq. (30) contains two parts.
One part is determined by the particular solution (26,31)
of the singular integral equation (20) 
with $K(W,W',Q^2)=0$.
It is, in fact, the particular solution of the integral equation (20).
Obviously, there are an infinite number
of particular solutions of Eq. (20) which differ
from each other by the solutions of homogeneous part of this
equation. The solution (26,31) and the particular solution
of Eqs. (20,30) generated by (26,31)
are concrete solutions which are entirely
determined by the Born term
$M^B(W,Q^2)$
and turns out to be 0 when
$M^B(W,Q^2)=0$. Let us denote this particular
solution of Eq. (20) as $M_{Born}^{part}(W,Q^2)$.

Other part of the solution 
of Eq. (20) is generated by 
$M^{hom}_{K=0}(W,Q^2)$. It is the 
solution of the homogeneous part
of Eq. (20), i.e. of Eq. (20) with 
$M^{B}(W,Q^2)=0$.
This part has an arbitrary weight, which can be
found only from some additional  conditions,
for example, from comparison with experiment.

The solutions 
$M_{Born}^{part}(W,Q^2)$ and
$M^{hom}(W,Q^2)$
can be found from the Fredholm
integral equation (30) only by numerical methods.
The results of the numerical calculations
for the multipoles
$M_{1+}^{3/2},E_{1+}^{3/2},S_{1+}^{3/2}$
at $Q^2=0, 1, 2, 3~GeV^2$ are presented
on Figs. 2-5. The particular solutions on Figs. 2-4
are normalized by the form factor corresponding
to the dipole formula:
\begin{equation}
G_D (Q^2)=1/(1+Q^2/0.71~{\rm GeV^2})^2.
\end{equation}
For the multipole
$S_{1+}^{3/2}$ 
the results 
are presented for the ratio
$S_{1+}^{3/2}/|\bf{k}|$,
as this multipole enter the cross section in the form
$S_{1+}^{3/2}/|\bf{k}|$ 
(see Appendix, Eqs. (A.3,A.4)).

At $Q^2=0$ the solutions of the integral equation (30)
for the multipoles 
$M_{1+}^{3/2},E_{1+}^{3/2}$
have been obtained in Refs. [6,7] too.
They coincide with our results.
For the multipole
$M_{1+}^{3/2}$
there is a slight, practically invisible,
difference in the particular solutions
which is caused by the fact that in Refs. [6,7]
high energy contributions to the dispersion
relations are taken into account.
The coincidence of our particular
solutions with those of Refs. [6,7]
which, in fact, takes place confirms our
statement, that high energy
contributions do not affect 
the particular solutions for
the considered multipoles.

The solutions of the homogeneous equation
are presented on Fig.5. They are normalized
on the same value at the energy 
$E_L\equiv (W^2-m^2)/2m=0.34~GeV$ 
corresponding to the center of the $P_{33}(1232)$ resonance.

Let us discuss now the interpretation of the obtained
solutions. Suppose, one describes the amplitudes
$M_{1+}^{3/2},E_{1+}^{3/2},S_{1+}^{3/2}$
within some dynamical model
in terms of contributions of the diagrams of Fig.1
taking into account rescattering effects.
The examples of such models can be found in Refs.[13,19].
In such approach the rescattering effects
produce imaginary part in the Born term which is by itself real.
The imaginary part produced by the final state interaction
in the diagrams of Fig.1(a-c) should be considered
as nonresonance background to the $P_{33}$
resonance contribution in the multipole amplitudes
$M_{1+}^{3/2},E_{1+}^{3/2},S_{1+}^{3/2}$.
For the first time this was mentioned
in Ref. [13]. The attempts to calculate
this nonresonance contribution within
dynamical models are connected with uncertainties
coming from the cutoff procedure and the method of taking
into account off-shell effects.
In dispersion relation approach
due to the elasticity of the $h_{1+}^{3/2}$
amplitude of $\pi N$ scattering
up to quite large energies and, as a result,
due to the validity of the Watson theorem
up to energies which are much larger in comparison
with the energies in the $P_{33}$ resonance region,
there is the possibility to find the nonresonance
contribution in the model independent way.
As is seen from Eq. (26) the contribution
produced by the final state interaction
in the Born term caused by the resonance
$\pi N$ scattering is reproduced
by dispersion relations in the form of particular
solution of the integral equation (20)
generated by the Born term.
The whole 
particular solution satisfies the requirements
of unitarity and crossing symmetry.
In Refs. [6,7] it is shown that the factor $D(W)$
in Eq. (26) is responsible for modification
of amplitudes at small distances. 
By the shape and magnitude the nonresonance 
contributions
into the multipoles
$M_{1+}^{3/2}$ and $E_{1+}^{3/2}$
at $Q^2=0$ 
obtained in this work in the form of
particular solutions of Eq. (20)
are very close to those 
obtained within dynamical model of Ref. [19].

The contribution of the diagram of Fig.1d together with rescattering effects
should be identified with the solution
of the homogeneous part of Eq. (20).
The rescattering effects modify the vertices
in this diagram. As a result, in the center of the
$P_{33}(1232)$
resonance the vertices $\gamma^* N P_{33}(1232)$
and $\pi N P_{33}(1232)$
should be considered as dressed vertices.
The dressed vertex $\pi N P_{33}$
can be found from experimental data on the width
of the $P_{33}\rightarrow \pi N$ decay.
Let us note that as is seen from Fig. 5 
the shapes of the homogeneous solutions,
i.e. the shapes of the $P_{33}$ contribution,
are slightly different for different multipoles
and for different values of $Q^2$.
The presence of such difference is natural
due to rescattering effects
which can be different
for different multipoles and at different $Q^2$.
For the comparison with predictions of quark model, 
QCD and other models the magnitudes of the
$\gamma^*N\rightarrow P_{33}(1232)$
form factors extracted from multipoles
at $W=W_r=1.232~ GeV$ should be used.

For the analysis
of pion electroproduction data a
phenomenological approach proposed by Walker [20]
is widely used. In this approach multipole amplitudes,
including the amplitudes
$M_{1+}^{3/2},E_{1+}^{3/2},S_{1+}^{3/2}$,
are parameterized in terms of resonances
taken in the Breit-Wigner form
and smooth background contribution. The
Breit-Wigner formula for the multipoles
$M_{1+}^{3/2},E_{1+}^{3/2},S_{1+}^{3/2}|{\bf k}_r|/|{\bf k}|$
used by Walker has the form:
\begin{equation}
M_{B-W}(W)=\frac{W_r {\bf\Gamma}}{W_r^2-W^2-iW_r{\bf \Gamma}}
\left(\frac{\bf q_r}{\bf q}\right)^2 \frac {|\bf k|}{\bar{|\bf k}_r|},
\end{equation}
where
\begin{equation}
{\bf\Gamma}=\Gamma\left(\frac {|\bf q|}{|\bf q_r|}\right)^3
\frac{{\bf q}_r^2+X^2}{{\bf q}^2+X^2},
\end{equation}
$\Gamma=0.114~ GeV,~X=0.167 ~GeV,~\bf q_r$
and $\bar{\bf k}_r$ are the momenta of pion
and real photon in c.m.s. in the center of the $P_{33}(1232)$
resonance.

Dispersion relations 
allow to check the Walker approach
for the multipoles 
$M_{1+}^{3/2},E_{1+}^{3/2}$, 
$S_{1+}^{3/2}$
in the $P_{33}(1232)$ resonance region.
As is seen from the obtained results
the nonresonance background in these
multipoles has nontrivial behavior and can not be
described by a smooth function. The shape of
the $P_{33}(1232)$ resonance contribution is
fixed in dispersion relation approach
by the solution of the homogeneous part of Eq. (20).
As is seen from Fig. 5 it also differs from that of 
the Breit-Wigner formula. So, the input for the multipoles
$M_{1+}^{3/2},E_{1+}^{3/2},S_{1+}^{3/2}$
obtained in this work within dispersion relation
approach does not coincide with that in the phenomenological
approach proposed by Walker. 
\section{Comparison with experiment
at $Q^2=0$.
Amplitudes $A_{1/2}^p,~A_{3/2}^p$ and $E2/M1$ ratio for the
$\gamma N\rightarrow P_{33}(1232)$ 
transition extracted from experiment.}
In this 
Section we will present our 
results on the description of experimental
data for the multipole amplitudes
$M_{1+}^{3/2}$ and $E_{1+}^{3/2}$
at $Q^2=0$ which are evaluated with high
accuracy from existing experimental data
in the multipole analysis of Ref. [21].
In the considered approach these data
should be described as a sum of the particular 
and homogeneous solutions
of the integral equations for
the multipoles
$M_{1+}^{3/2}$ and $E_{1+}^{3/2}$
obtained in the previous Section.
The particular solutions have definite magnitude
fixed by the Born term.
The weights of the homogeneous solutions 
should be found from the requirement 
of the best description
of the experimental data. For this aim we have used
fitting procedure. The obtained results
together with the experimental data are presented
on Figs. 6,7. In order to demonstrate
the role of the nonresonance background
contributions on these figures separately 
the particular solutions generated by
the Born term are given. 
The homogeneous solutions taking with the weights obtained
in the result of fitting the experimental data
are also presented.
The obtained homogeneous solutions
give the following values of multipoles
$M_{1+}^{3/2}$ and $E_{1+}^{3/2}$ corresponding
to the contribution of the $P_{33}(1232)$
resonance:
\begin{eqnarray}
&&M_{1+}^{3/2}(\gamma N\rightarrow P_{33}(1232))
=4.22\pm 0.12~\mu b^{1/2}, \\
&&E_{1+}^{3/2}(\gamma N\rightarrow P_{33}(1232))
=-0.055\pm 0.011~\mu b^{1/2}. 
\end{eqnarray}

In Table 1 we present the ratio 
$E_{1+}/M_{1+}$ for the transition
$\gamma N\rightarrow P_{33}(1232)$ 
which follows from (40,41)
and the helicity amplitudes for this transition
obtained by the following formulas:
\begin{eqnarray}
&&A_{1+}^{3/2}=-
\left[\frac{3}{8\pi}\frac{|\bf k|}{|\bf q|}\frac{m}{M}
\frac{\Gamma_\pi}{\Gamma^2}\right]^{1/2}A^p_{1/2}, \\
&&B_{1+}^{3/2}=
\left[\frac{1}{2\pi}\frac{|\bf k|}{|\bf q|}\frac{m}{M}
\frac{\Gamma_\pi}{\Gamma^2}\right]^{1/2}A^p_{3/2}, 
\end{eqnarray}
where $A_{1+}=(M_{1+}+3E_{1+})/2,~
B_{1+}=E_{1+}-M_{1+}$ and $M,\Gamma,~\Gamma_\pi$
are the 
$P_{33}(1232)$
mass, total and partial widths.
In Table 1 we present also the ranges of the 
amplitudes $A_{1/2}^p,~A_{3/2}^p$
from Ref. [17] which are extracted
from existing experimental data without taking into
account nonresonance contribution
in the multipole $M_{1+}^{3/2}$.
The magnitudes of these amplitudes desagree with quark model
predictions. As it is seen from our results
this desagreement is removed due
to taking into account the nonresonance background
produced by rescattering effects in the diagrams
corresponding to the Born term. 
\section {Conclusion}
The set of the following features of the
multipole amplitudes $M_{1+}^{3/2},E_{1+}^{3/2},S_{1+}^{3/2}$
in the $P_{33}(1232)$ resonance region:

a. the elasticity of the corresponding amplitude $h_{1+}^{3/2}$
of $\pi N$ scattering up to energies which are much higher 
than the energies in the 
$P_{33}$ resonance region and, as a result,
the validity of the Watson theorem in this energy region with the known
phase $\delta _{1+}^{3/2}(W)$;

b. the smallness of high energy contributions
and other multipole contributions to the dispersion
relations for the multipoles
$M_{1+}^{3/2},E_{1+}^{3/2},S_{1+}^{3/2}$,
\newline
allows to transform the dispersion relations for
these multipoles into the integral equations
of the Fredholm type, where the terms which are responsible
for the inhomogeneouty of these equations
are determined only by the Born terms. 
This allowed us to make strict conclusions on the nonresonance contributions
into the multipoles $M_{1+}^{3/2},E_{1+}^{3/2},S_{1+}^{3/2}$
and on the shape of the resonance contributions which are equivalent 
to the solutions of the homogeneous parts of the integral equations.
The weights of resonance contributions are not fixed
by dispersion relations. These are the only unknown 
parameters in the multipoles $M_{1+}^{3/2},E_{1+}^{3/2},S_{1+}^{3/2}$
which should be found from experimental data.
\newline
\begin{center}
{\large {\bf {Acknowledgments}}}
\end{center}

I am grateful to V. Burkert, N. Isgur, B. Mecking and A.V. Radyushkin
for useful discussions and to S.G. Stepanyan for help in numerical calculations. 
I express my gratitude for the hospitality
at Jefferson Lab where the final part of this work was accomplished.
The work was supported in part by Armenian Foundation of
Scientific Researches (Grant \# 94-681)
and by INTAS Grant \# 93-283 ext.  
 \vspace{1cm}

{\Large \bf Appendix}

\vspace{0.3cm}

\renewcommand\theequation{A.\arabic{equation}}

\setcounter{equation} 0

The residues in Eq. (8) are equal to:
\begin{eqnarray}
&&R_1^{(v,s)}=\frac{ge}{2}(F_1^{(v,s)}+2mF_2^{(v,s)}),\nonumber \\
&&R_2^{(v,s)}=-\frac{ge}{2}F_1^{(v,s)}(Q^2),\nonumber \\
&&R_3^{(v,s)}=-\frac{ge}{4}F_1^{(v,s)}(Q^2), \\
&&R_5'^{(v,s)}=\frac{ge}{4}(\mu -Q^2-t)F_2^{(v,s)}(Q^2),\nonumber \\
&&R_6^{(v,s)}=ge F_2^{(v,s)}(Q^2),\nonumber \\
&&R_8^{(v,s)}=\frac{ge}{2}F_2^{(v,s)}(Q^2),\nonumber
\end{eqnarray}
where in accordance with existing experimental data we have:
\begin{eqnarray}
&&e^2/4\pi=1/137,~~g^2/4\pi=14.5,\nonumber\\
&&F_1^{(v,s)}=\left( 1+\frac{g^{(v,s)}\tau}{1+\tau}\right)
G_D(Q^2),\nonumber \\
&&F_2^{(v,s)}=\frac{g^{(v,s)}}{2m}\, \frac{G_D(Q^2)}{1+\tau},\\
&&F_\pi (Q^2)=1/(1+Q^2/0.59~{\rm GeV^2}),\nonumber \\
&&\tau =Q^2/4m^2, \quad g^{(v)}=3.7, \quad g^{(s)}=-0.12.\nonumber
\end{eqnarray}
Amplitudes $f_i$ introduced in Sec. 2 are related
to the multipole and helicity amplitudes
and to the cross section in the following way:
\begin{eqnarray}
&&f_1=\sum\left\{(lM_{l+}+E_{l+})P'_{l+1}(x)+\left[(l+1)M_{l-}+E_{l-} \right]
P'_{l-1}(x)\right\},\nonumber \\
&&f_2=\sum\left[(l+1)M_{l+}+lM_{l-} \right] P'_l(x),\nonumber \\
&&f_3=\sum\left[(E_{l+}-M_{l+})P''_{l+1}(x)+(E_{l-}+M_{l-})P''_{l-1}(x)\right],\\
&&f_4=\sum(M_{l+}-E_{l+}-M_{l-}-E_{l-})P''_l(x),\nonumber \\
&&f_5=\sum\left[(l+1)S_{l+}P'_{l+1}(x)-lS_{l-}P'_{l-1}(x)\right],\nonumber \\
&&f_6=\sum\left[lS_{l-}-(l+1)S_{l+}\right] P'_l(x),\nonumber
\end{eqnarray}

\begin{eqnarray}
&&H_1\equiv f_{+,+-}=f_{-,-+}=-\cos\frac{\theta }{2} \sin \theta
(f_3+f_4)/\sqrt{2},\nonumber \\
&&H_2\equiv f_{-,++}=-f_{+,--} =-\sqrt{2} \cos\frac{\theta }{2}
\left[f_1-f_2- \sin^2\frac{\theta }{2}(f_3-f_4)\right],\nonumber\\
&&H_3\equiv f_{+,-+}=f_{-,+-}=\sin\frac{\theta }{2} \sin \theta
(f_3-f_4)/\sqrt{2},\\
&&H_4\equiv f_{+,++}=f_{-,--} =\sqrt{2} \sin\frac{\theta }{2}
\left[f_1+f_2+ \cos^2\frac{\theta }{2}(f_3+f_4)\right]
,\nonumber \\
&&H_5 \equiv f_{-,-0}=-f_{+,+0}=-\frac{Q}{|{\bf k}|} \cos
\frac{\theta }{2}(f_5+f_6),\nonumber \\
&&H_6 \equiv f_{+,-0}=f_{-,+0}=\frac{Q}{|{\bf k}|} \sin
\frac{\theta }{2}(f_5-f_6),\nonumber
\end{eqnarray}

\begin{eqnarray}
&&\frac{\bar{k}}{|\bf q|} \frac{d \sigma }{d \Omega_\pi}=\frac{1}{2}
\left(|H_1|^2+ |H_2|^2 +|H_3|^2+|H_4|^2
\right)+\nonumber \\
&&+\varepsilon \left(|H_5|^2+ |H_6|^2 \right)-\varepsilon \cos 2\varphi
Re\left(H_4 H_1^*-H_3 H_2^* \right)- \\
&&-\cos \varphi \left[ \varepsilon(1+\varepsilon)\right]^{1/2}Re
\left[H_5^*(H_1-H_4)+H_6^*(H_2+H_3)\right],\nonumber
\end{eqnarray}
where $x=cos\theta$, $\theta$ and
$\varphi$ are the polar and azimuthal
angles of the pion in c.m.s., $\bf k$ and
$\bf q$ are the momenta of the photon and the pion
in this system, $\varepsilon$ is the polarization
factor of the virtual photon,
$f_{\mu_2,\mu_1\lambda}$ are the helicity
amplitudes, $\lambda,~\mu_1,~\mu_2$
are the helicities of photon, initial and
final nucleons, respectively, $\bar{k}=(W^2-m^2)/2W$.

\newpage
{\Large \bf {Figure Captions}}
\vspace{1cm}
\newline
{\large \bf{Fig. 1}} Diagrams corresponding to the contributions
of the Born term (nucleon and pion poles) and the $P_{33}(1232)$
resonance to the production of pions on nucleons
by virtual photons.
\newline
{\large \bf{Fig. 2}} The multipole amplitude $M_{1+}^{3/2}$.
 The imaginary parts of the partial
solutions of the dispersion relations (20) for this amplitude
at different $Q^2$
generated by the Born term. 
The solutions are divided by the dipole form factor (37).
These solutions should be considered as background
contributions (to the $P_{33}(1232)$ resonance)
produced by rescattering effects in the diagrams 
corresponding to the Born term. $E_L\equiv \frac{W^2-m^2}{2m}$.
\newline
{\large \bf{Fig. 3}} The multipole amplitude $E_{1+}^{3/2}$.
The legend is as for Fig. 2. 
\newline
{\large \bf{Fig. 4}} The multipole amplitude $S_{1+}^{3/2}/|{\bf k}|$.
The legend is as for Fig. 2. 
\newline
{\large \bf{Fig. 5}} The imaginary parts of the 
solutions of the dispersion relations (20)
with $M^B(W,Q^2)$=0 for
the multipoles $M_{1+}^{3/2},~E_{1+}^{3/2},~S_{1+}^{3/2}$
at different $Q^2$. These solutions represent
the shape of the $P_{33}(1232)$ resonance
contribution into multipoles.
The curves 1-4 correspond
to the following multipoles:
(1) $S_{1+}^{3/2}|{\bf k}_r|/|{\bf k}|$ at $Q^2=0$;
(2) $E_{1+}^{3/2}$ at $Q^2=0$;
(3) $M_{1+}^{3/2}$ at $Q^2=0$ and
$E_{1+}^{3/2}$, $S_{1+}^{3/2}|{\bf k}_r|/|{\bf k}|$ at $Q^2=1-3 ~GeV^2$;
(4) $M_{1+}^{3/2}$ at $Q^2=1-3 ~GeV^2$.
For comparison the shapes of these multipoles
corresponding to the Breit-Wigner formula (38)
are presented: dotted line - $Q^2=0$,
dashed line - $Q^2=1-3 ~GeV^2$.
\newline
{\large \bf{Fig. 6}} The multipole amplitude $M_{1+}^{3/2}$
at $Q^2=0$.
 Our results for 
the imaginary part of this amplitude (solid line)
in comparison with experimental data [21].
Separately, the nonresonance background contribution
given by the particular solution of eq. (20)
(dashed line) and the resonance contribution
given by the solution of the homogeneous part
of this equation (dotted line) are presented.
\newline
{\large \bf{Fig. 7}} The multipole amplitude $E_{1+}^{3/2}$
at $Q^2=0$. The legend is as for Fig. 6.
\newpage
{\Large \bf {Table Captions}}
\vspace{1cm}
\newline
{\large \bf{Table 1}} Helicity amplitudes and $E2/M1$
ratio for the $\gamma N\rightarrow P_{33}(1232)$
transition. Our results are extracted from the
phase shift analysis data
of Ref. [21] taking into account the 
nonresonance contributions into multipoles
given by the particular solutions 
of the dispersion relations generated
by the Born term.
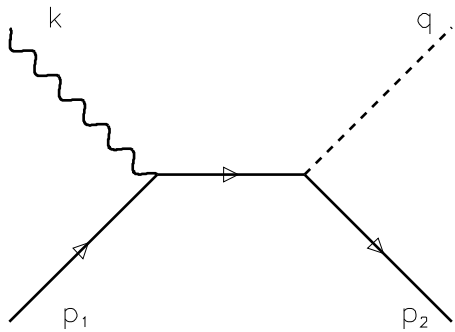
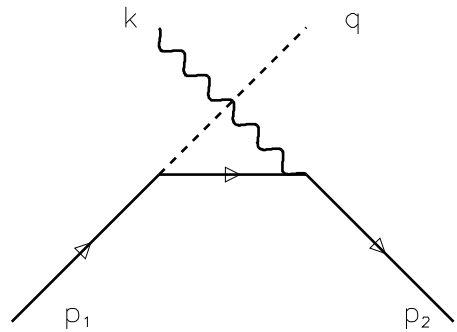
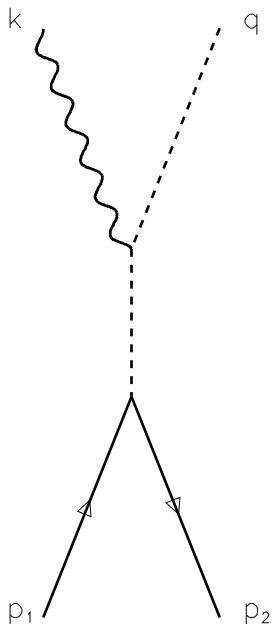
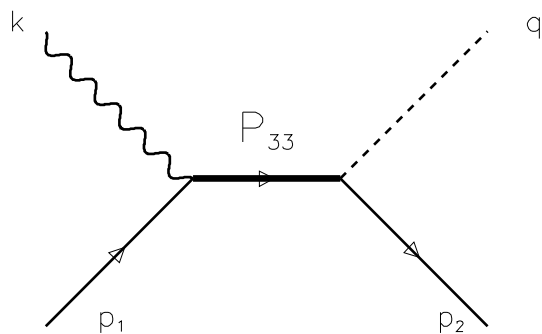
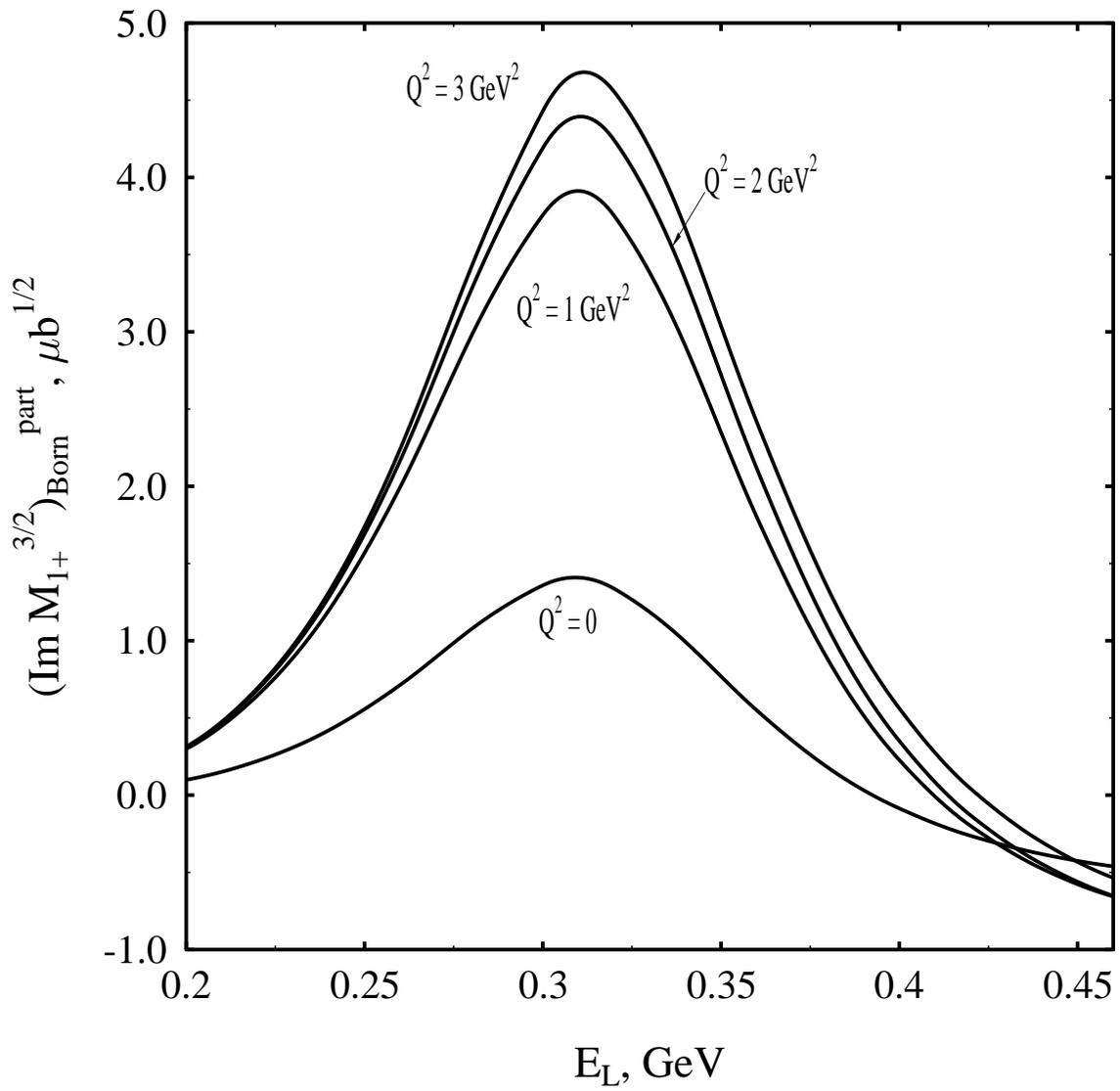
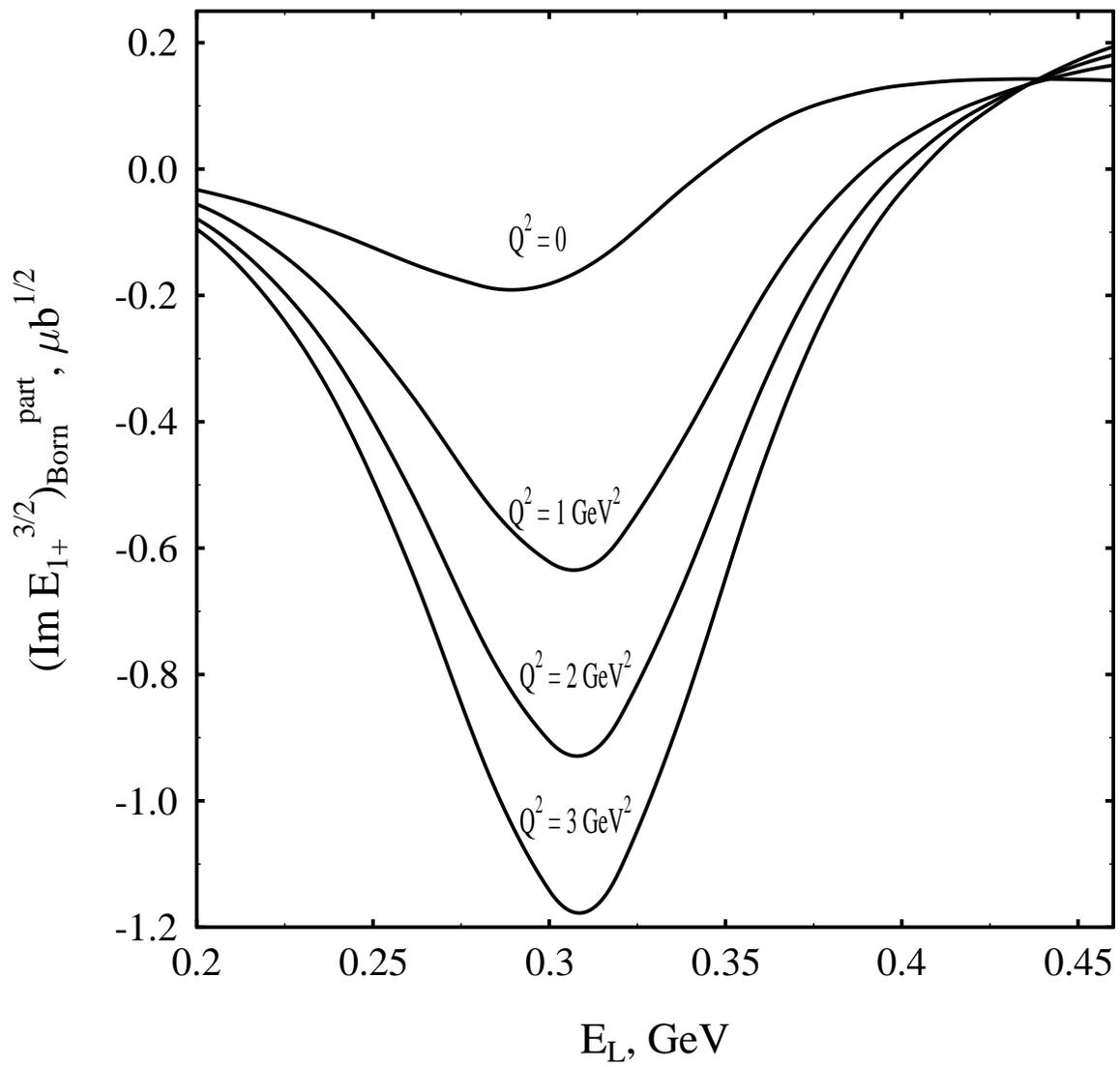
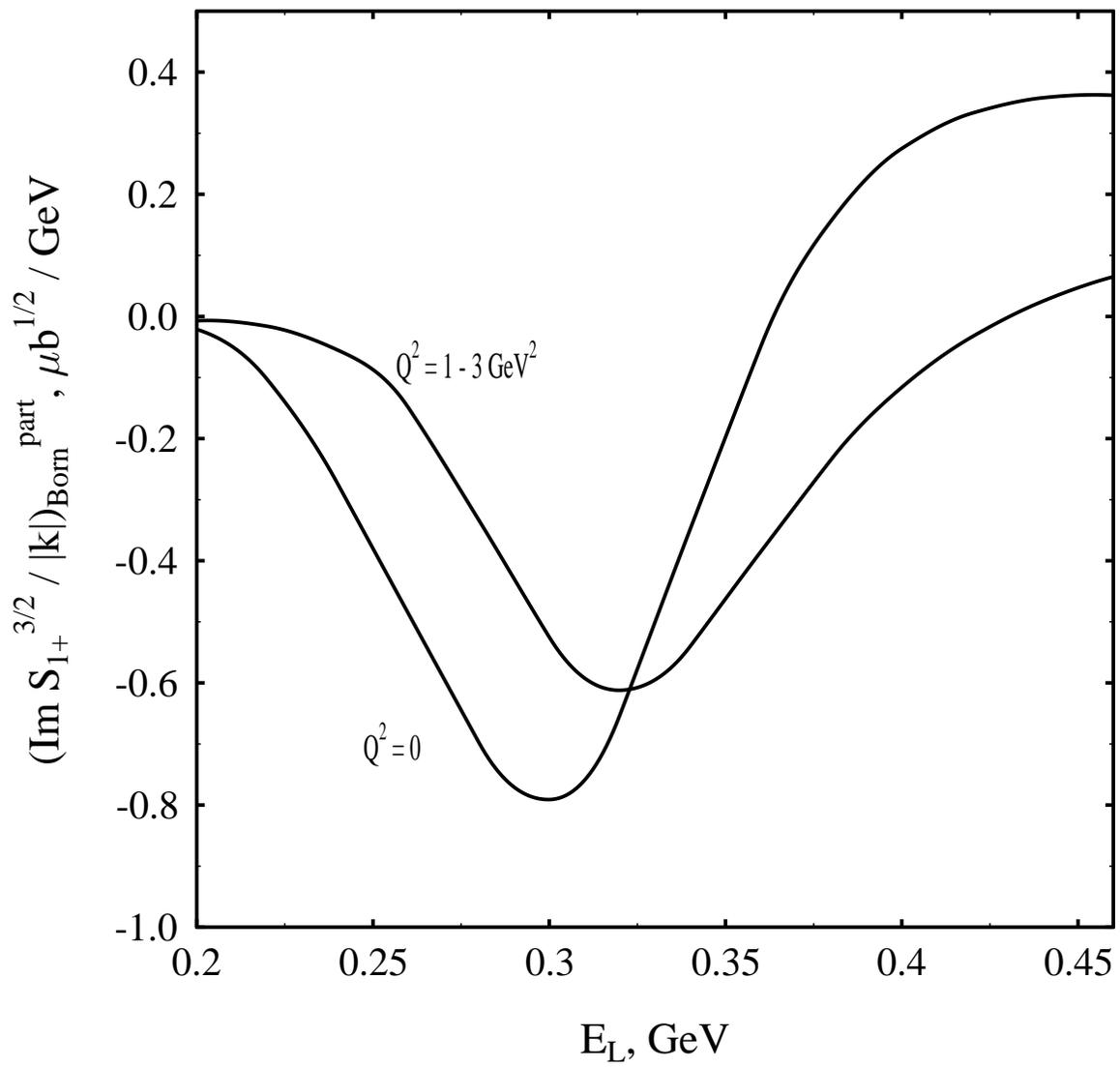
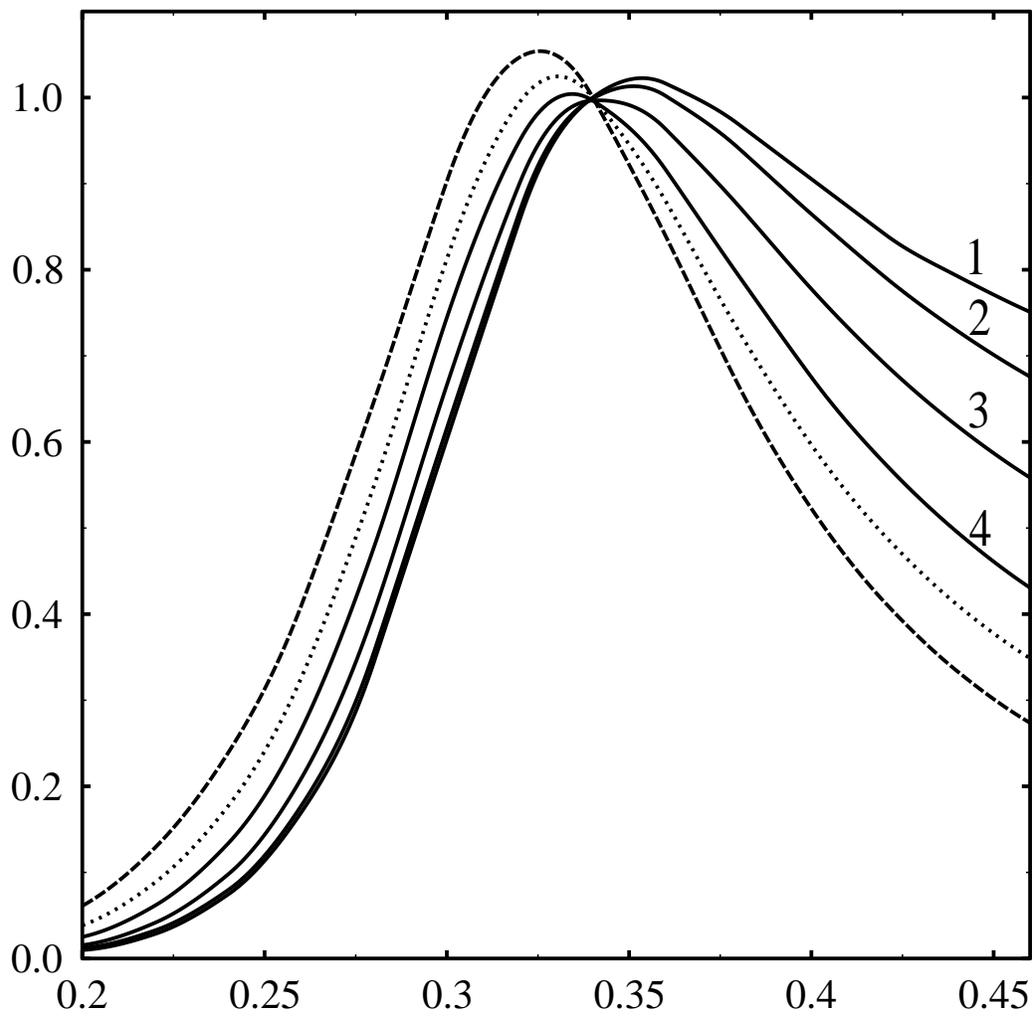
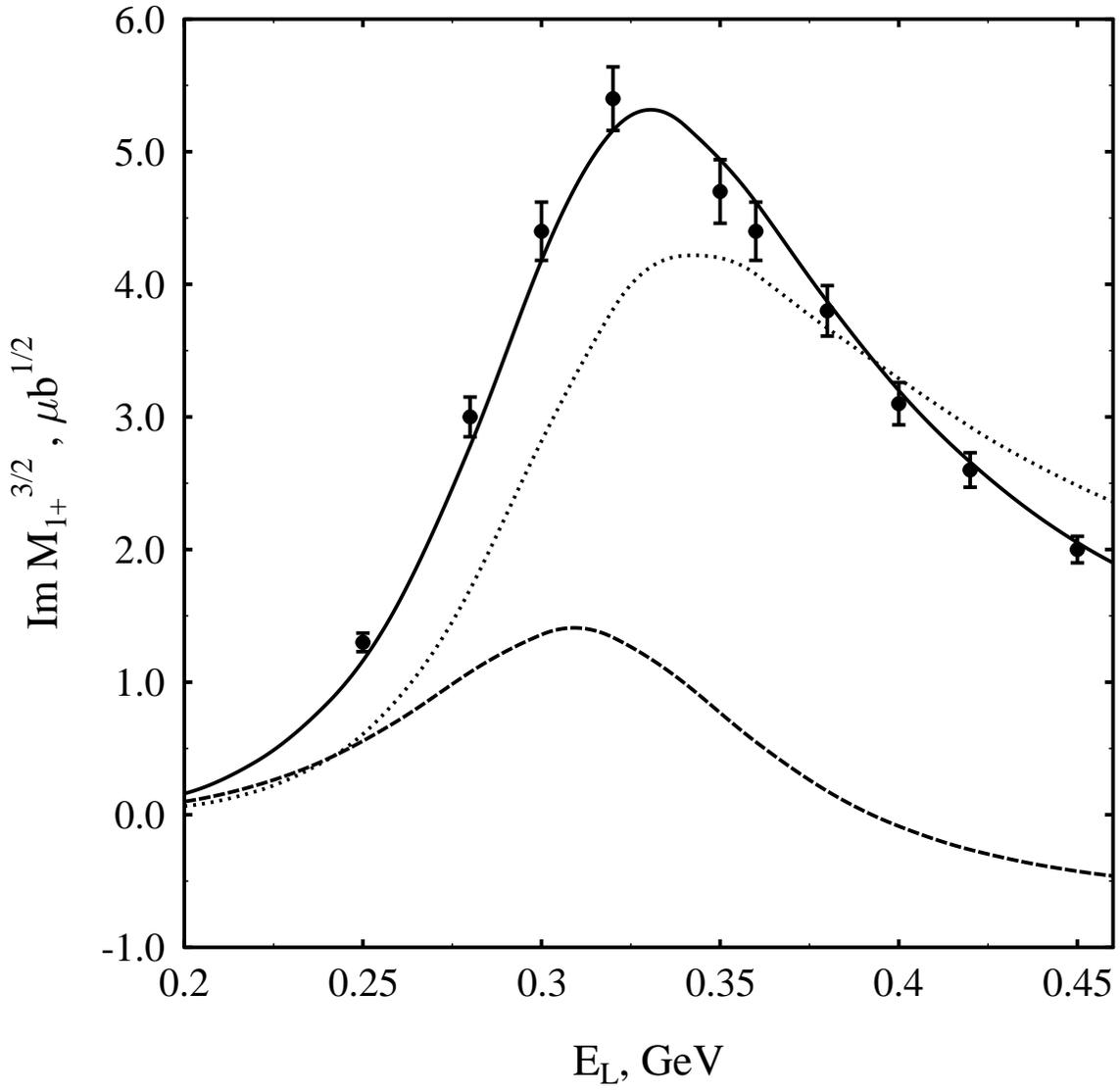
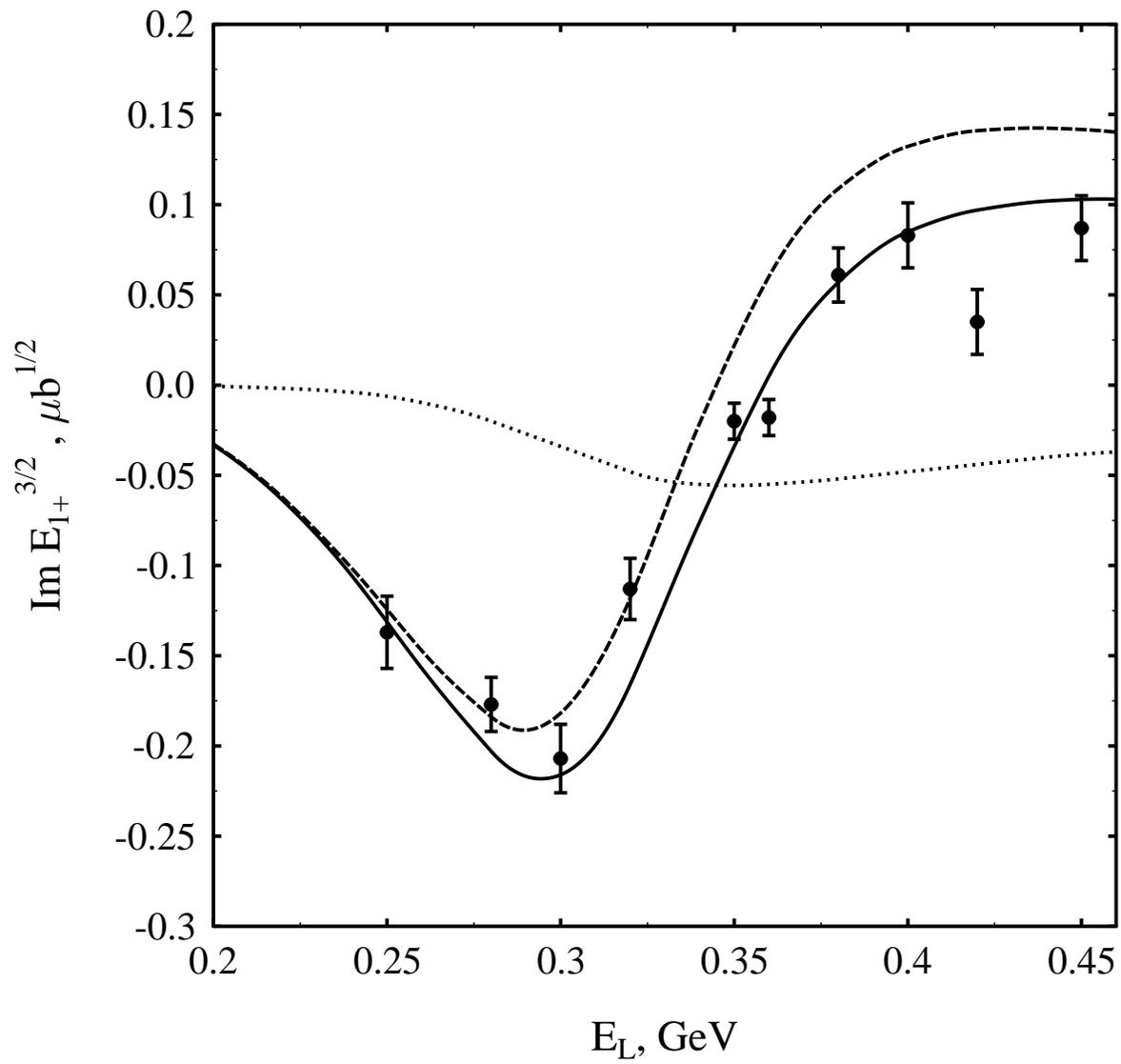
\end{document}